\documentclass[iop,apj]{emulateapj}
\usepackage[breaklinks,colorlinks,citecolor=blue]{hyperref}

\shorttitle{EUV Waves in $^{3}$He-Rich SEP Events}
\shortauthors{R.~Bu\v{c}\'ik et al.}

\begin{document}


\submitted{Draft version September 6, 2015}

\title{Observations of EUV Waves in $^{3}$He-Rich Solar
Energetic Particle Events}


\author{R.~Bu\v{c}\'ik and D.~E.~Innes and L.~Guo}
\affil{Max-Planck-Institut f\"{u}r Sonnensystemforschung, D-37077 G\"{o}ttingen, Germany}
\affil{Max Planck/Princeton Center for Plasma Physics, Princeton, NJ 08540, USA}
\email{bucik@mps.mpg.de}

\author{G.~M.~Mason}
\affil{Applied Physics Laboratory, Johns Hopkins University, Laurel, MD 20723, USA}

\and

\author{M.~E.~Wiedenbeck}
\affil{Jet Propulsion Laboratory, California Institute of Technology, Pasadena, CA 91109, USA}



\begin{abstract}
Small $^{3}$He-rich solar energetic particle (SEP) events with their anomalous
abundances, markedly different from solar system, provide evidence for
a unique acceleration mechanism that operates routinely near solar active
regions. Although the events are sometimes accompanied by coronal mass
ejections (CMEs) it is believed that mass and isotopic fractionation is
produced directly in the flare sites on the Sun. We report on a large-scale extreme ultraviolet (EUV) coronal wave
observed in association with $^{3}$He-rich SEP events.
In the two examples discussed, the observed waves were triggered by minor flares
and appeared concurrently with EUV jets and type III radio bursts but without
CMEs. The energy spectra from one event are consistent with so-called class-1
(characterized by power laws) while the other with class-2
(characterized by rounded $^{3}$He and Fe spectra) $^{3}$He-rich SEP events,
suggesting different acceleration mechanisms in the two. The observation of EUV
waves suggests that large-scale disturbances, in addition to
more commonly associated jets, may be responsible for the production
of $^{3}$He-rich SEP events.
\end{abstract}





\section{INTRODUCTION}

Discovered more than 40 years ago, $^{3}$He-rich solar energetic particles (SEPs) are still
poorly understood. The enormous abundance enhancement (up to factors of $>$10$^{4}$)
of the rare $^{3}$He isotope is the most striking feature of these events, though
large enhancements in heavy (Ne-Fe) and ultra-heavy nuclei are also observed
\citep[see][for a review]{koc84,mas07}. Type III radio bursts
with their parent low-energy electrons are firmly associated with $^{3}$He-rich SEP
events \citep{rea85,nit06}. Solar sources of $^{3}$He-rich SEPs
are often accompanied by jet-like emissions in extreme ultraviolet (EUV) or
X-ray images and sometimes in white-light reaching out to distances $>$2 solar
radii ($R_{\sun}$) \citep{wan06,nit06,nit08}. Somewhat surprisingly
an association with fast and narrow coronal mass ejections (CMEs) has been
reported in some $^{3}$He-rich SEP events \citep{kah01,nit06}. CMEs with
their driven shocks have been considered as a particle source for large
gradual events while $^{3}$He-rich SEPs are believed to be produced directly
in flare sites presumably through gyroresonant wave-particle interactions
\citep[see review by][]{rea13}. The CMEs associated with $^{3}$He-rich
SEP events may be considered to be high-altitude counterparts
of reconnection X-ray jets discovered by \citet{shi92}.

What other signatures of solar activity can be associated with $^{3}$He-rich SEPs?
Recently \citet{nit15} describe the properties of a large sample (29) of $^{3}$He-rich
events with SDO observations of their solar source regions. About a half
of the events were associated with jets and another half
with wider eruptions. Four were associated with large scale disturbances
(EUV waves) and slow ($<$350~km\,s$^{-1}$) CMEs. Coronal
EUV waves in large gradual SEP events have been observed for almost two
decades \citep{tor99}. Some authors have recently attempted to explain SEP events observed at
widely separate longitudes in terms of EUV waves \citep[e.g.,][]{rou12,par13,lar14}.
At the outset of the eruption they appear to trace the progress of the shock
responsible for particle acceleration. However, the association of EUV waves
with $^{3}$He-rich SEP events has seldom been discussed \citep{wie13,nit15}.


In this paper, we report on two $^{3}$He-rich SEP events clearly associated
with coronal EUV waves. The observations were made during a period of
low solar activity in early 2010. We examine the relationship of the EUV waves to
the accompanying energetic particles, including a detailed analysis of the wave
fronts. It appears that the wave properties may be correlated with the observed
ion spectra. These observations of SEP events and EUV waves are presented in section~\ref{obs},
and we discuss their implications in section~\ref{sad}.

\section{OBSERVATIONS} \label{obs}


The two $^{3}$He-rich SEP events reported in this paper were identified using
observations from the time-of-flight mass spectrometers Ultra Low Energy
Isotope Spectrometer \citep[ULEIS;][]{mas98} on the {\sl Advanced Composition
Explorer} ({\sl ACE}) and the Suprathermal Ion Telescope \citep[SIT;][]{mas08} on
the {\sl Solar Terrestrial Relations Observatory AHEAD} spacecraft ({\sl STEREO-A}). The
responsible solar sources were examined in full-Sun images from the EUV
imager \citep[EUVI;][]{how08} on {\sl STEREO}. {\sl ACE} is in an orbit around the L1 point;
{\sl STEREO-A} is in a heliocentric orbit at $\sim$1~AU near the ecliptic plane
moving faster than Earth at a rate of $\sim$22\degr\ per year.

\subsection{$^{3}$He-Rich SEP Events} \label{ev}

\begin{figure}
\plotone{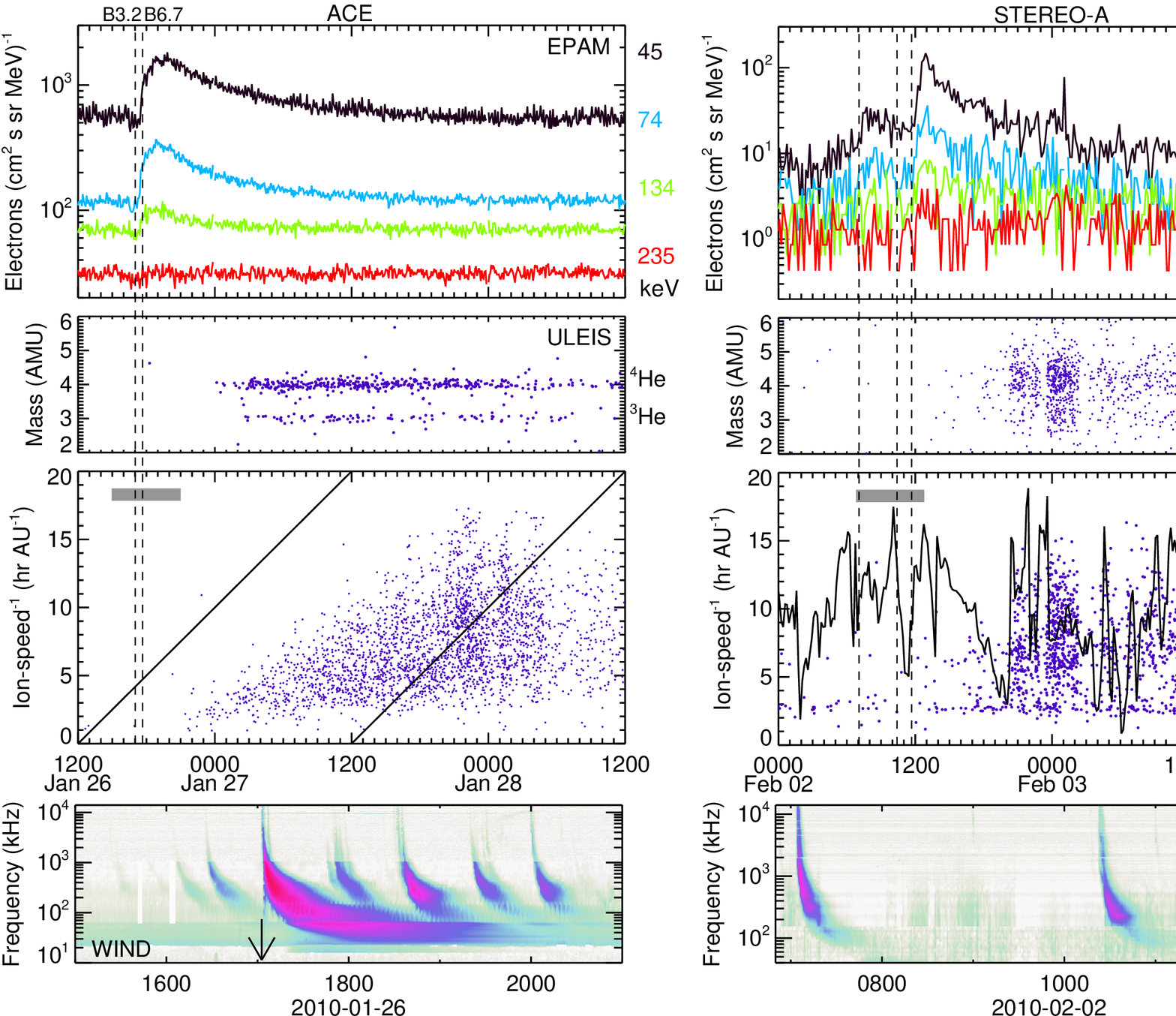}
\caption{(a) 5 minute EPAM/{\sl ACE} (left) and 10 minute SEPT/{\sl STEREO-A} (right)
electron intensities. (b) ULEIS/{\sl ACE} 0.4-10~MeV\,nucleon$^{-1}$ (left) and
SIT/{\sl STEREO-A} 0.25-0.90~MeV\,nucleon$^{-1}$ (right) He mass spectrograms. (c) ULEIS
(left) and SIT (right) spectrograms of 1/ion-speed versus arrival times of
10-70~amu ions. Sloped lines (left) indicate arrival times for particles
traveling along a field line of 1.2~AU without scattering. Black curve (right)
is the magnetic field zenith angle in Radial-Tangential-Normal (RTN)
coordinates. The dashed vertical lines mark the start times of the {\sl GOES}
X-ray flares (left) and type III bursts (right). (d) radio spectrogram from
{\sl WIND} (left) and {\sl STEREO-A} (right) WAVES instruments during 6~hr periods
indicated by horizontal shaded bars in panels (c). The arrows mark type III bursts
associated with the events. \label{fig1}}
\end{figure}

\begin{figure}
\plotone{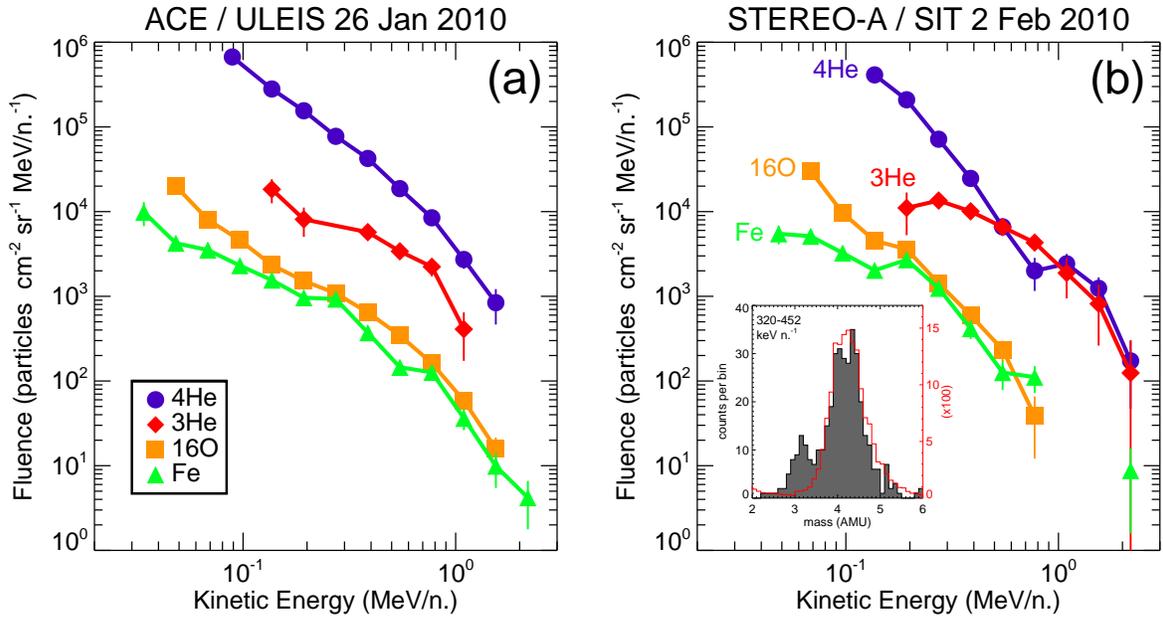}
\caption{Energy spectra for January 26 (a) and February 2 (b) $^{3}$He-rich SEP
events. Gray shaded histogram is for February 2 event. Red histogram is for
corotating interaction region events (2010 May-June) to compare SIT observations
with no $^{3}$He mass peak. \label{fig2}}
\end{figure}

Figure~\ref{fig1} shows two $^{3}$He-rich SEP events, one observed by {\sl ACE} on 2010 January 26
(left panels) and another by {\sl STEREO-A} on 2010 February 2 (right). These are
among the first $^{3}$He-rich SEP events detected after the unusually long solar
minimum between solar cycles 23 and 24 (the February 2 event is also
identified in \citet{wie13}). Both events were accompanied by
energetic electron enhancements observed by EPAM/{\sl ACE} \citep{gol98} or
SEPT/{\sl STEREO-A} \citep{mul08} as shown in Figure~\ref{fig1}a. The
approximate onset of EPAM electron event on January 26 is at 17:25~UT (45~keV). The onset of
SEPT electron event on February 2 is quite uncertain; the main increase starts around
12:00~UT (50~keV), but it was preceded by another minor event. Figure~\ref{fig1}b shows individual ions in the helium mass
range at 0.4-10~MeV\,nucleon$^{-1}$ for ULEIS/{\sl ACE} and at 0.25-0.90~MeV\,nucleon$^{-1}$
for SIT/{\sl STEREO-A}. The $^{3}$He track is clearly separated in ULEIS
data. An examination of SIT mass histograms in the February 2 event confirms
a clear $^{3}$He peak within the helium range (see Figure~\ref{fig2}b). The January 26
event has a clear dispersive onset, where higher energy ions arrived earlier
than lower energy ones as shown by the triangular pattern in the inverted
ion-speed time spectrogram in Figure~\ref{fig1}c. The February 2 event could
also have been dispersive, but it is possible that a magnetic-cloud like
structure may have reduced the intensity of energetic ions \citep[e.g.,][]{can06}.
The interplanetary magnetic field (IMF) data from the MAG/{\sl ACE}
instrument \citep{acu08} shows a smooth rotation of the magnetic field vector
during 14-20~UT (see Figure~\ref{fig1}c).
The 320-450~keV\,nucleon$^{-1}$ $^{3}$He/$^{4}$He ratio is 0.13 for January 26 and 0.41 for
February 2 event. The Fe/O ratio is $\sim$0.6 in both events; somewhat smaller
than the average value in $^{3}$He-rich flares \citep[cf. $\sim$0.95 at
385~keV\,nucleon$^{-1}$ in][]{mas04}.

The January 26 event was associated with a B3.2 {\sl GOES} X-ray flare in active
region (AR) 1042 (N20\degr W75\degr) with a start time at 17:01~UT (Solar
Events List\footnote{\url{www.swpc.noaa.gov}}).
Close to the time of the flare onset, a strong type III radio burst was observed by
WAVES/{\sl WIND} \citep{bou95} (see Figure~\ref{fig1}d). The estimated ion solar
release time from extrapolation of ULEIS spectrogram data to the zero
propagation time is around the type III burst onset, though uncertainty
arising from this technique has been reported to be $\pm$45 minutes
\citep{mas00}. Less clear is the ion release time in the February 2
event. The event (from different AR; see Section~\ref{wav2}) may be related to the
preceding main electron increase and associated type III burst at 11:42~UT
measured by WAVES/{\sl STEREO-A} \citep{bou08}; other type III bursts
preceding the event occurred at 10:24 and 07:04~UT (see Figure~\ref{fig1}d). The
burst at 07:04~UT is also significant, extending to both high and low
frequencies, and may be associated with an observed minor electron intensity
increase. Hereafter, we focus on the type III burst at 11:42~UT because it was associated
with the main electron event. The Solar Radio Bursts report\footnote{\url{ftp.ngdc.noaa.gov}}
includes type III at 17:03 on January 26 in frequency range 25-144~MHz
(Sagamore Hill) and type III at 07:04 on February 2 in 20-130~MHz (Culgoora).
The type III at 11:42~UT on February 2 extended into the range
20-70~MHz\footnote{\url{secchirh.obspm.fr}} (Nan\c{c}ay Decameter array).
No metric type II radio bursts were observed in these two events.

Figure~\ref{fig2}a shows event averaged fluence spectra for January 26 and
Figure~\ref{fig2}b shows the spectra for the February 2 $^{3}$He-rich SEP event. The
January 26 spectra for $^{3}$He, $^{4}$He, O, and Fe have similar power laws. They are
reminiscent of class-1 event spectra where major species exhibit similar power
laws or broken power laws with $^{3}$He often showing stronger hardening below
$\sim$1~MeV\,nucleon$^{-1}$ \citep{mas00,mas02}. In the February 2 event,
$^{4}$He and O have similar power laws but the $^{3}$He and Fe spectra are distinctly
flatter, leading to a larger variation of $^{3}$He/$^{4}$He and Fe/O with energy
than in the January 26 event. The February 2 event spectra are
similar to class-2 event spectra, characterized by curved
$^{3}$He and Fe spectra towards low energies with $^{3}$He rollovers in the range
$\sim$100-600~keV\,nucleon$^{-1}$ and Fe rollovers below $\sim$100~keV\,nucleon$^{-1}$ \citep{mas00}.
Certainly, the spectral shapes in these two events are not very representative
of class-1 or class-2 events. There are fluctuations at several spectral
points making it somewhat difficult to categorize the January 26 event solely
by its $^{3}$He shape. Note, the median value of $^{3}$He/$^{4}$He ratio of
class-1 events \citep[$\sim$0.12 at 385~keV\,nucleon$^{-1}$ in][]{mas02} is
strikingly similar to $^{3}$He/$^{4}$He ratio in the January 26 event.

\subsection{EUV Wave - January 26 Event} \label{wav1}

\begin{figure}
\plotone{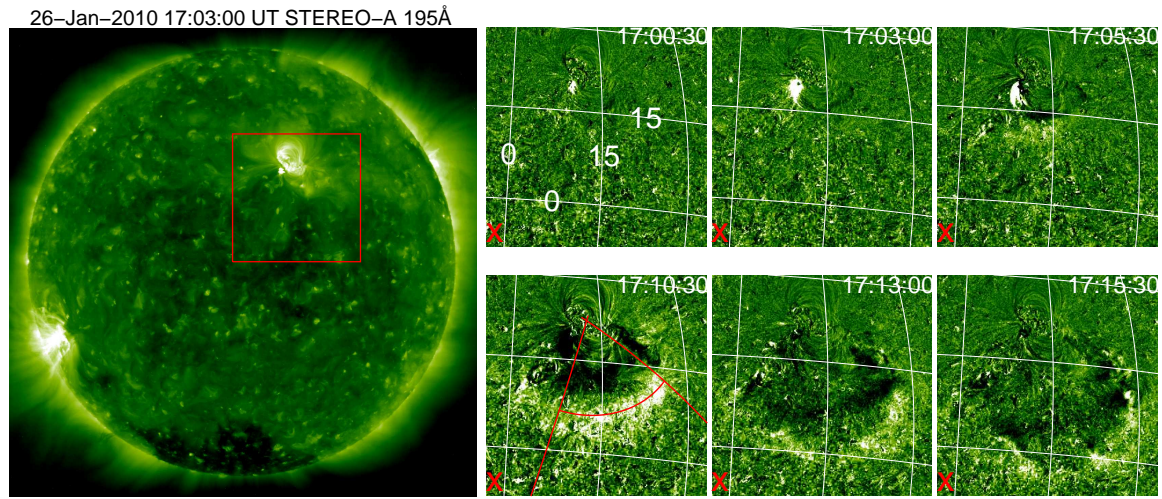}
\caption{(Left) {\sl STEREO-A} 195~{\AA} EUV image of the solar disk on 2010 January 26
17:03~UT. (Right) six panels - 2.5 minute (except for 17:10:30, which is 5 min)
running difference images of the area around AR 1042 marked by red square in
the direct image. Red crosses indicate nominal foot-point of IMF line
connecting to L1. Two red curves, passing through the AR (panel 17:10:30), indicate
a 70\degr\ arc sector on the solar sphere where the EUV intensity profiles in
Figure~\ref{fig5} were determined. The red curve along the wave front outlines
an arc centered on the AR. \newline (An animation of this figure is available.) \label{fig3}}
\end{figure}

During the investigated events the {\sl ACE} and {\sl STEREO-A} were angularly separated
by 65\degr\ allowing the near-west limb regions from {\sl ACE} to be observed in a more
direct view by {\sl STEREO-A}. We emphasize that such a constellation enables a
completely new insight on $^{3}$He-rich sources, not available in earlier
investigations. Figure~\ref{fig3}~(left) shows the EUV 195~{\AA} image of AR 1042 near the
central meridian in the {\sl STEREO-A} view. Running difference images in
Figure~\ref{fig3}~(right) show jet-like emissions at 17:00-17:05~UT in the eastern
foot-point of a series of small magnetic loops. The observed temporal
coincidence between the EUV jet in AR 1042 and the type III burst indicates
that the AR contains open field allowing particles to escape.

A large-scale wave was observed emanating from AR 1042,
the $^{3}$He-rich SEP source, as demonstrated in running difference images in
Figure~\ref{fig3}. The wave was clearly seen even in direct EUV images (see animation of Figure~\ref{fig3}~(left)). The launch
time of the wave temporally coincides with the EUV jet. A bright wave front
was clearly seen at 17:05~UT ($\sim$4 minutes after the X-ray flare and 2 minutes
after the type III) but a weaker arc-shaped dimming, probably associated with
the wave, was already seen at the type III burst onset. The wave front
propagated southward towards the equator and nominal spacecraft foot-point
based on the Parker spiral model. The nominal foot-point location was not
reached before 17:15~UT, but the electron, and likely also the ion, release
was associated with the jet, type III burst and the wave-launch that occurred
earlier. Thus the wave could already intersect the open field lines connecting to the
spacecraft at the time when it started.

The nose of the wave front traveled 15\degr\ in latitude between 17:03 and 17:13~UT,
which corresponds roughly to 300~km\,s$^{-1}$. This is within the range
(200-400~km\,s$^{-1}$) of typical EUV wave speeds \citep{tho09} and is
comparable with quiet-Sun fast magnetosonic speeds. Previous observations
have shown that EUV waves can be faster in the early stage and may even be
shocks \citep[e.g.,][]{war11}. Indeed, newer high-cadence observations,
capable of capturing the initial phases of the wave evolution, indicate much
higher speeds \citep[$\sim$600~km\,s$^{-1}$;][]{nit13} implying that these waves may
steepen to shocks quite frequently. EUV wave shocks directly observed as
dome-like enhancements propagating ahead of a CME have been reported in some
recent investigations \citep{ver10,koz11,ma11}.

The EUV image of the Sun's disk in Figure~\ref{fig3} reveals quite uniform coronal
structure with the only other AR located far away, near the south-east solar
limb and a coronal hole at the south-pole. This likely enabled undisturbed
wave propagation and therefore its easier observation. Note that ARs/coronal
holes in the paths of the waves usually cause them to fade/reflect
\citep[e.g.,][]{tho99}. The fact that AR 1042 is quite isolated with
no simultaneous activity observed in other regions also allowed a more
straightforward identification of the $^{3}$He-rich SEP source. Note that many
$^{3}$He-rich SEP events have been left without identified solar sources, for
example, 40$\%$ out of 117 events by \citet{nit06}.

The driver of EUV waves has often been associated with CMEs but in this event
there was no CME. Any CME associated with January 26 event would be best visible in a coronograph
from the Earth view because of the source location near the western limb.
The {\sl SOHO} LASCO C2 observations, covering the range 1.5-6~$R_{\sun}$, show a narrow
stream at 17:54, 18:06 and 18:30~UT at near-equatorial region on the west.
In {\sl SOHO} LASCO CME catalog\footnote{\url{cdaw.gsfc.nasa.gov/CME\_list}}
this brief eruption was classified as a very poor event. No eruption was seen
from 1.4 to 4 $R_{\sun}$ in {\sl STEREO-A} COR-1 5-minute running difference
images\footnote{\url{cdaw.gsfc.nasa.gov/stereo/daily\_movies}}. Also
the {\sl STEREO} COR1 preliminary list\footnote{\url{cor1.gsfc.nasa.gov/catalog}}
indicates no associated CME.

\subsection{EUV Wave - February 2 Event} \label{wav2}

\begin{figure}
\plotone{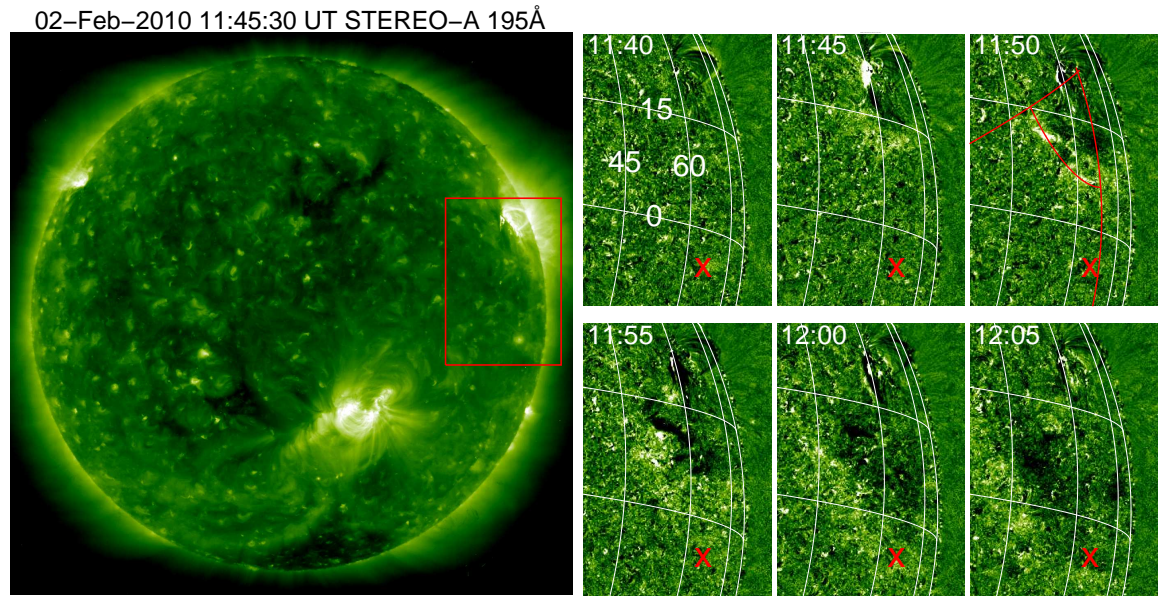}
\caption{(Left) STEREO-A 195~{\AA} EUV image of the solar disk on 2010 February 2
11:45~UT. (Right) six panels - 5 minute running difference images of the area
marked by red square in the direct image. Red crosses indicate foot-point for
STEREO-A. The red curves in the panel 11:50 have the same meaning as in
Figure~\ref{fig3}. \newline (An animation of this figure is available.) \label{fig4}}
\end{figure}

Figure~\ref{fig4}~(left) shows the {\sl STEREO-A} EUV 195~{\AA} image of the solar source AR for
the February 2 $^{3}$He-rich SEP event, located at N20\degr W65\degr\ from
{\sl STEREO-A}, which was 65\degr\ west of the Sun-Earth line. Running
difference images in Figure~\ref{fig4}~(right) show a jet at 11:45~UT, temporally
coincident with a type III burst shown in Figure~\ref{fig1}d. Similar to the previous
event, the jet was emitted from the Sun's surface at the eastern foot-point
of a series of small-scale loops. The AR emerged on 2010 January 30 when it
was at the west limb as seen from the Earth. The running difference images
in Figure~\ref{fig4} show a bright wave front emitted from the AR around the time of
the jet and propagating in the south-east direction. The wave reached
{\sl STEREO-A} nominal foot-point around 12~UT, but the electron (and likely the ion also)
release occurred earlier at 11:45~UT in association with the jet and the type
III radio burst. The wave front in the February 2 event appears to be more
diffuse (cf. wave fronts 12-13~min after type III burst onsets in both
events at 17:15 and 11:55~UT) and less bright than in the January 26 event
(cf. wave fronts 7-8~min after type III burst onsets in both events at
17:10 and 11:50~UT). The projection effects in the February 2 event probably play a minor role as the
wave propagates toward the central meridian where these effects are less
dominant. Because of the diffuse fronts in the February 2 event it is more
difficult to measure the wave speed. A rough estimate is $\gtrsim$200~km\,s$^{-1}$ between
11:45 and 12:00~UT where the wave front traveled 15\degr\ in the latitude.
In direct EUV images the wave was not so clearly seen (see animation of Figure~\ref{fig4}~(left)) as in the
previous event. Note that in the February 2 event, {\sl STEREO-A} provides only 5
minute cadence images while in January 26 event the cadence is 2.5 minutes.
Thus, insufficient temporal resolution may be one reason why these waves were
not noticed in earlier $^{3}$He-rich SEPs investigations. The above mentioned
projection effects due to western location of $^{3}$He-rich sources may also add
to the difficulty in an identification of the associated waves. The less
intense type III burst at 10:25~UT (see Figure~\ref{fig1}d) was also associated with
a jet and a coronal wave but these were less significant than in the 11:42~UT
type III burst. The type III at 07:04~UT also coincided with EUV brightening
in the same AR but no wave was observed. The {\sl STEREO-A} COR1 running
difference images showed from 12:15~UT onward a small bright feature moving
outward towards the west. This weak outflow was not marked in the COR-1
Preliminary CME List.

\subsection{EUV Wave Profiles} \label{wav3}

\begin{figure}
\epsscale{.6}
\plotone{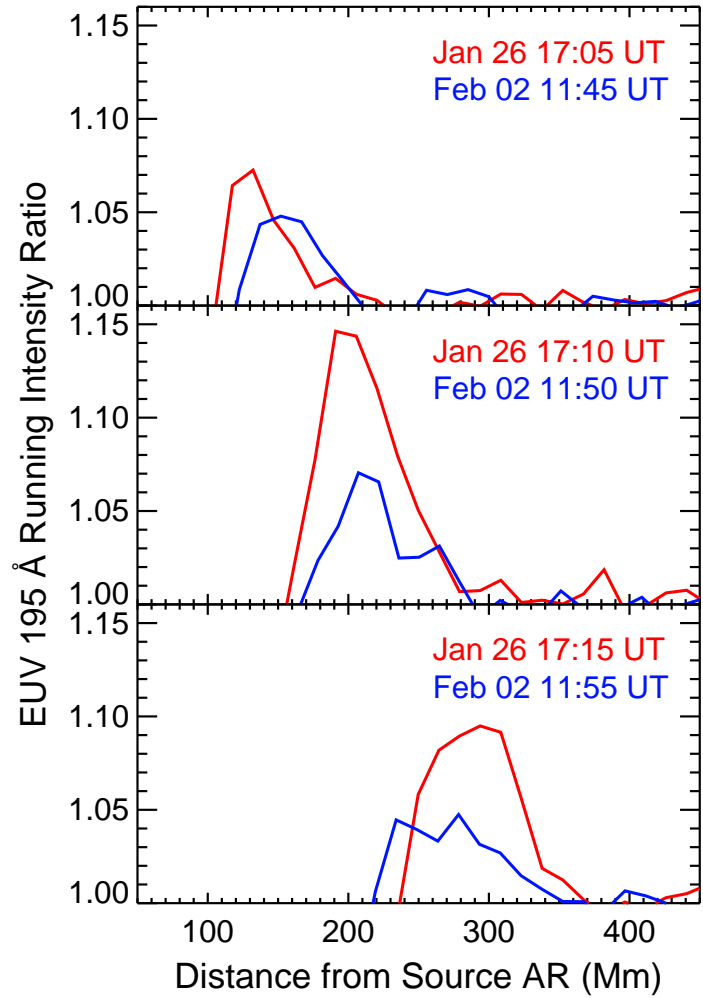}
\caption{Average EUV intensity-ratio profiles for the January 26 (red curve)
and February 2 (blue curve) wave fronts at three different times. The average ratios
were determined in a 70\degr\ annular sectors with radial width 15~Mm and center
in the source AR. The 70\degr\ arc sectors were placed on the brightest portion
of the wave fronts (see Figures~\ref{fig3} and \ref{fig4}). \label{fig5}}
\end{figure}

In addition to a visual inspection as given in the previous sections,
we also provide a quantitative analysis
of the wave fronts. Figure~\ref{fig5} shows the evolution of the wave front profiles within 12 and 13 minutes
after the associated type III radio bursts in the January 26 and Februay 2 events,
respectively. A similar approach where the intensity ratios are derived
along the propagating fronts is presented in earlier studies
\citep[e.g.,][]{ver10}. The figure reveals that the amplitudes of
the intensity ratio and their temporal fluctuations are larger for the January 26 wave.
The temporal behavior of the trailing front edges suggests that the January 26 wave
was likely accelerating while the February 2 wave was moving with more uniform speed.
These profiles also indicate lower speed for the February 2 wave.




\section{DISCUSSION AND SUMMARY} \label{sad}

This paper examines a new solar phenomenon observed in association with
$^{3}$He-rich SEPs. We present two events, where in addition to EUV jets, $^{3}$He-rich
source ARs launched simultaneously a coronal EUV wave. The waves were
initiated by minor flares and not linked to CMEs. The EUV waves have been
considered to be closely related with CMEs \citep{bie02}, though
recent investigations \citep{nit13} indicate that the association is
not so strong.

The EUV waves may be a more common feature in $^{3}$He-rich SEP events than previously
thought. \citet{wie13} have noticed large-scale disturbances in
the source region of the STEREO-B 2010 February 7 and ACE February 8 $^{3}$He-rich
SEP events. They have suggested that EUV waves could be responsible for a sympathetic
flaring in a region far from the nominal connection and thus contributing to
a wide particle longitude distribution. In a parallel study, \citet{nit15} have
reported large scale propagating fronts in a few $^{3}$He-rich SEP events on ACE
by examining also active periods of a current solar cycle. Note that some other
$^{3}$He-rich SEP events like 2008 November 4 on ACE \citep{mas09} or 2011 July 1
on STEREO-B \citep{buc14} are also associated with EUV waves (not mentioned
in original studies). Their sources are well visible near the central meridian
similarly to the 2010 January 26 event.

Coronal jets, a signature of magnetic reconnection between magnetic loops and
overlying open field \citep{shi92}, create turbulence for $^{3}$He-rich SEP
acceleration \citep[e.g.,][]{mil98,pet12}. Nearly all models of $^{3}$He-rich
SEPs require some kind of two-stage processes. For example electrons,
accelerated by cascading turbulence, excite plasma waves \citep{mil98} capable
of gyroresonant interaction with ambient $^{3}$He \citep{tem92}.
An alternative to turbulence is the reconnection-exhaust ion heating followed
by an interaction with multiple magnetic islands \citep[e.g.,][]{dra12}.
The two spectral classes of $^{3}$He-rich SEPs is a relatively new feature, and
have not yet been adequately explained. It has been suggested that class-2
events represent the basic mechanism of $^{3}$He enrichment and that
class-1 events need a further stage of acceleration (e.g. by a shock wave)
that may modify the spectra \citep{mas02}. A recent
study by \citet{nit15} indicated a slight tendency for events with jets at their
source to have class-2 spectra (9 of 13), and those with EUV waves at their
source to have class-1 spectra (3 of 4).

The observation of EUV waves, which may be shocks in their early phases, calls
into question the details of $^{3}$He-rich SEP acceleration. The shape of the energy
spectra observed in the first event, when the wave was bright, is similar to
class-1 events. In this example the wave might have been capable of modifying
the original curved spectra created by turbulence or by the pick-up mechanism
in the reconnection exhausts. \citet{bie02} noted that EUV waves with
bright, sharp fronts may indicate shocks. In the second event, where the wave
front was less bright and perhaps slower, the spectral forms were consistent with the class-2
events, suggesting that the influence of the wave was minor. The energy
spectra of $^{3}$He and $^{4}$He in the February 2 event are quite similar to spectra
in the class-2 event on 1999 September 30 \citep{mas00}, which were
excellently fitted by stochastic acceleration  \citep[Figure~9 in][]{liu06}.
We need to examine more events in order to evaluate the relevance of EUV waves
on the two class spectra of $^{3}$He-rich SEPs. The two examples presented here
also raise the question whether EUV waves themselves (even not steepened into
the shocks) may generate or enhance turbulence required for $^{3}$He-rich SEP
acceleration models.




\acknowledgments

We are grateful to the referee for valuable comments that helped to improve
the manuscript. This work was supported by the Max-Planck-Gesellschaft zur F\"{o}rderung der
Wissenschaften. The {\sl STEREO} SIT is supported by the Bundesministerium f\"{u}r Wirtschaft through
the Deutsches Zentrum f\"{u}r Luft- und Raumfahrt (DLR) under grant 50 OC 1301.
{\sl ACE}/ULEIS and {\sl STEREO}/SIT are supported at APL by NASA grant NNX13AR20G/115828
and NASA through subcontract SA4889-26309 from the University of California
Berkeley. The work at JPL and Caltech was supported through subcontract SA2715-26309
from UC Berkeley under NASA contract NAS5-03131T, and by NASA grants NNX11A075G
and NNX13AH66G.

\clearpage



\clearpage

\clearpage

\clearpage

\end{document}